\documentclass{vldb}
\usepackage[utf8]{inputenc}
\usepackage{color}
\usepackage{balance}  
\usepackage{graphicx}
\usepackage{subcaption}

\newcommand{\tuple}[1]{\left<#1\right>}
\newcommand{\comprehension}[2]{\left\{\;#1\;|\;#2\;\right\}}
\newcommand{\tinysection}[1]{\smallskip \noindent \textbf{#1}.~~}
\newtheorem{example}{Example}

\numberofauthors{3} 

\author{
\alignauthor
Poonam Kumari\\
       \affaddr{University at Buffalo}\\
       \email{poonamku@buffalo.edu}
\alignauthor
Said Achmiz\\
       \email{achmizs@gmail.com}
\alignauthor
Oliver Kennedy\\
       \affaddr{University at Buffalo}\\
       \email{okennedy@buffalo.edu}
}

\title{Communicating Data Quality in On-Demand Curation}
\date{}

\begin{document}

\maketitle

\begin{abstract}
On-demand curation (ODC) tools like Paygo, KATARA, and Mimir allow users to defer expensive curation effort until it is necessary.
In contrast to classical databases that do not respond to queries over potentially erroneous data, ODC systems instead answer with guesses or approximations.  
The quality and scope of these guesses may vary and it is critical that an ODC system be able to communicate this information to an end-user.  
The central contribution of this paper is a preliminary user study evaluating the cognitive burden and expressiveness of four representations of ``attribute-level'' uncertainty.  
The study shows (1) insignificant differences in time taken for users to interpret the four types of uncertainty tested, and (2) that different presentations of uncertainty change the way people interpret and react to data.
Ultimately, we show that a set of UI design guidelines and best practices for conveying uncertainty will be necessary for ODC tools to be effective.  This paper represents the first step towards establishing such guidelines.
\end{abstract}

\section{Introduction}
Historically, the quality of a dataset would be ensured before it was analyzed, often through complex, carefully developed curation processes designed to completely shield analysts from any and all uncertainty.  This curation establishes trust in the data, which in turn helps to establish trust in the results of analyses.  
However, as typical data sizes and rates grow, this type of brute-force \textit{upfront} curation process is becoming increasingly impractical.  As a result, analysts have started turning to new, ``on-demand'' or ``pay-as-you-go'' approaches~\cite{Beskales2014,Dallachiesa:2013:NCD:2463676.2465327,Jeffery:2008:PUF:1376616.1376701,Khayyat:2015:BSB:2723372.2747646,Roy:2015:EQA:2856318.2856329,Yang:2015:LOA:2824032.2824055} to data curation.  
On-demand curation (ODC) systems minimize the amount of upfront time and effort required to load, curate, and integrate data. 
Data stored in an ODC is, initially at least, of low quality and queries are liable to produce incomplete or incorrect results.  
To mitigate the unreliability of these results, ODC systems typically provide a form of provenance or lineage, tracking the effects of uncertainty through queries and tagging results with relevant quality metrics (e.g., confidence bounds, standard deviations, or probabilities). 
If the analyst finds the result quality insufficient, the ODC can help her to prioritize her data curation efforts.

Most ODC efforts are specialized forms of
\textit{probabilistic databases}~\cite{suciu2011probabilistic} that allow for queries over uncertain, probabilistically defined data.
Classical probabilistic databases produce outputs either in the form of ``certain'' answers (that provide only limited practical utility), or in the form of probability distributions.  
Representing a query output as a distribution alleviates the monotonous (and error-prone) task of handling probabilities, error conditions, and outliers in the middle of a query.
Nevertheless, error-handling logic is still necessary, even if it is never expressly declared; 
A human interpreting the results must decide whether and how to act on the results given.
Just having a probability distribution for query results is insufficient: \textit{the uncertainty must be communicated to the users who will ultimately act on the results}.
Complicating matters further is the fact that many database users lack the extensive background in statistics necessary to interpret complex probability distributions.  

In this paper, we present our initial efforts to explore how probabilistic databases can communicate uncertainty about query results to their users.  
Fundamentally, we are interested in how the database should represent potential errors in tabular data being presented to the user.
A representation that communicates too much information can create an unnecessary cognitive burden for users.
Conversely, if a representation communicates too little, the user may not realize that data values are compromised and act on invalid information.  

To explore this tradeoff between imposed cognitive burden and efficacy, we conducted a preliminary user study with 14 participants drawn from the Department of Computer Science and Engineering at the University at Buffalo.  
We explored four different representations of one specific form of data uncertainty called attribute-level uncertainty.
Our results show that the choice of how to communicate low-quality data has a substantial impact on how users react to that information.
Responses to different representations ranged from a desire for more information, an efficient use of presented contextual details, and even included mild fear responses to the data being presented.
Thus, we argue that the design of interface elements for representing uncertainty is a critical part of probabilistic databases, ODCs, and data quality research in general.  
Concretely, this paper makes the following contributions:
(1) We outline a user study that explores four different presentations of attribute-level uncertainty.
(2) We quantitatively analyze the tradeoff between cognitive burden and decision-making based on results from our study.
(3) We qualitatively analyze the different representations' effects on study participants' thought processes.

\section{Background}

A probabilistic database~\cite{suciu2011probabilistic} $\tuple{\mathbb D, P}$ is typically defined as a set of deterministic database instances $D \in \mathbb D$ that share a common schema, and a probability measure $P : \mathbb D \mapsto [0,1]$ over this set.  Under \textit{possible worlds semantics}, a deterministic query $Q$ may be evaluated on a probabilistic database by (conceptually) evaluating it simultaneously on all instances in $\mathbb D$, producing a set of relation instances: 
$$Q(\mathbb D) = \comprehension{Q(D)}{D \in \mathbb D}$$
Note that these semantics also induce a probability measure over the set of possible query results as a marginal of $P$ computed over the result set.

Numerous semi-automated tools for curating low-quality data~\cite{Beskales2014,Dallachiesa:2013:NCD:2463676.2465327,Yang:2015:LOA:2824032.2824055,DBLP:journals/corr/NandiYKGFLG16} emit probabilistic database relations.  These relations model the ambiguity that arises during automated data curation, most frequently appearing in one of three forms: (1) Row-level uncertainty, (2) Attribute-level uncertainty, and (3) Open-world uncertainty.  
Row-level uncertainty arises when a specific tuple's membership in a relation is unknown.
Attribute-level uncertainty arises when specific values in the database are not known precisely.
Finally, open-world uncertainty arises when a relation can not be bounded to a finite set of possible tuples.   

\begin{figure}
\centering
\begin{tabular}{c|c|c|c|c}
 & \multicolumn{3}{c|}{\textbf{Rating Source}}\\
\textbf{Product} & \textbf{Buybeast} & \textbf{Amazeo} & \textbf{Targe} & \textbf{Note} \\ \hline
Samesung & 4.5 & 3.0 & 3.5 \\
Magnetbox & 2.5 &    & 3.0 \\
Mapple & 5.0 & 3.5 & & {\tiny Not a TV?}
\end{tabular}
\caption{\textbf{Examples of uncertainty.}}
\label{fig:uncertainty}
\vspace*{-3mm}
\end{figure}

\begin{example}
The example spreadsheet given in Figure~\ref{fig:uncertainty} shows reviews for 3 fictional television products from 3 fictional sources.  Each of the three types of uncertainty are illustrated: It is unclear whether the Mapple is actually a television (row-level uncertainty).  There are ratings missing for both the Magnetbox and the Mapple (attribute-level uncertainty).  Finally, there is the possibility that the spreadsheet is incomplete and there are television products missing (open-world uncertainty).
\end{example}

Several mechanisms for presenting probability distributions to end-users have been proposed.  
A common approach is to present only so-called ``certain'' answers~\cite{Fagin:2005:DEG:1061318.1061323} --- the subset of the output relation with no row- or attribute-level uncertainty.
Although computing certain answers presents a computationally interesting challenge, completely excluding low-quality results significantly decreases the utility of the entire result set.
Another common approach is to compute statistical metrics like expectations or variances for attribute-level uncertainty, and per-row probabilities (confidences) for row-level uncertainty.  
Presenting this information to users in a way that can be clearly distinguished from deterministic data is challenging.  
Thus, systems like MayBMS~\cite{Huang:2009:MPD:1559845.1559984} and MCDB~\cite{jampani2008mcdb} typically require users to explicitly request specific statistical metrics as part of queries.
The mental overhead of manually tracking which attributes of a dataset are uncertain is an unnecessary burden on users; In multiple efforts where we have attempted to deploy probabilistic databases in practice~\cite{Kennedy:2011:JEO:1989323.1989410,Yang:2015:LOA:2824032.2824055,DBLP:journals/corr/NandiYKGFLG16}, manual management of uncertain data has proven to be a non-starter.


Uncertainty also arises in other contexts.  For example, Online Aggregation~\cite{Hellerstein:1997:OA:253262.253291} uses sampling to approximate and incrementally refine results for aggregate queries.  The user interface
explicitly gives an expectation, confidence bounds, and \% completion, clearly communicating that the result is an approximation, and the level of quality a user can expect from it.
A second example, Jigsaw~\cite{Kennedy:2011:JEO:1989323.1989410} simulates what-if scenarios, producing graphs that illustrate possible outcomes over time.  Uncertainty is presented visually, with error bars and secondary lines used to show standard-deviations.  
Wrangler~\cite{Kandel:2011:WIV:1978942.1979444} helps users to visualize errors in data: A ``data quality'' bar communicates the fraction of data in each column that conforms to the column's type and the number of blank records.
Finally, the Mimir system~\cite{Yang:2015:LOA:2824032.2824055,DBLP:journals/corr/NandiYKGFLG16} uses automatic data curation operators that tag curated records with markers that persist through queries.  These markers manifest as highlights that communicate the presence of attribute and row-level uncertainty.  Users click on fields or rows to learn more about why the value/row is uncertain.  

\section{Experimental Design}
The experiment consisted of a ranking task where participants were presented with a web form that had a 3x3 matrix showing three ratings each for three products.  Participants were told that the ratings came from three different sources and were normalized to a scale of 1 to 5, with 5 being best and 1 being worst.
Each participant was asked to evaluate the products for purchase by ranking the products in the order of their preference.
A total of 14 participants, predominantly students in the Department of Computer Science and Engineering at the University at Buffalo, participated in the experiment.  

To ensure a roughly predictable ordering from participants, ratings for each product were generated uniformly at random with the following constraints: 
Ratings for one of the three products (henceforth termed `A') relative to a second product (termed `B') had to include one extremely favorable comparison for A (one source gave A a rating at least 1 higher than B), one somewhat favorable comparison (one source gave A a rating at least as high as B but no more than 1 higher), and one disfavorable comparison (the final source gave A a rating worse than or equal to B but no more than 1 worse).  Similar comparisons also had to hold between B and the final product (C).  These constraints were designed to elicit a ranking of `A', `B', and `C' from participants deciding based on either majority vote or based on the average of the three ratings.

Participants were asked to complete either one or five rounds of survey, with each round consisting of four trials.  
A single trial consisted of a single ranking task.  The first \textbf{Certain} trial in each round served as a control: The matrix shown was generated exactly as described above.
The remaining trials in each round each evaluated a single representation of uncertainty.  
In these trials, base data generation followed an identical process.
However, in each trial, one of the following representations of uncertainty was used to annotate a small number (2-4) of product rating values. 
(1) \textbf{Asterisk}: Some ratings were marked with an asterisk (e.g., \texttt{4.5*})  and participants were informed that these values were uncertain.  
(2) \textbf{Colored text}: The text of some ratings was colored red (e.g., \textcolor{red}{4.5}) and participants were informed that these fields were uncertain.  
(3) \textbf{Confidence interval}: Some ratings were annotated with $\pm X$ where $X \in [0.5, 1.5]$ (e.g., \texttt{4.5 +/- 0.5}) and participants were informed that the value for those fields could range over the indicated interval.  
(4) \textbf{Color coding}: The cells containing some ratings were given a red background (e.g., \fcolorbox{black}{red}{\texttt{4.5}}) and participants were informed that these fields were uncertain.

Interactions with the web-form --- such as product selection, re-ordering the product list, and submitting the participant's final order --- were logged along with timestamps. In addition to interactions with the web form, the experiment also used a think-aloud protocol: Participants were asked to verbalize their thought process while performing the task.  Audio logs were transcribed and the anonymized transcriptions were tagged and coded for analysis.

\section{Efficiency and Effectiveness}
The two primary questions that we sought to answer for each of the four representations of uncertainty were (1) Is the representation \textit{effective} at communicating uncertainty, and (2) What is the \textit{cognitive burden} of interpreting the representation? 
Concretely, we identified at least three distinct behavioral responses to uncertainty in the data presented, suggesting differences in the efficacy of each representation.  
We also noted that all four representations of uncertainty required a similar amount of decision time, suggesting that all four representations impose similar cognitive burdens.  

\begin{figure}
\centering
\includegraphics[width=\columnwidth]{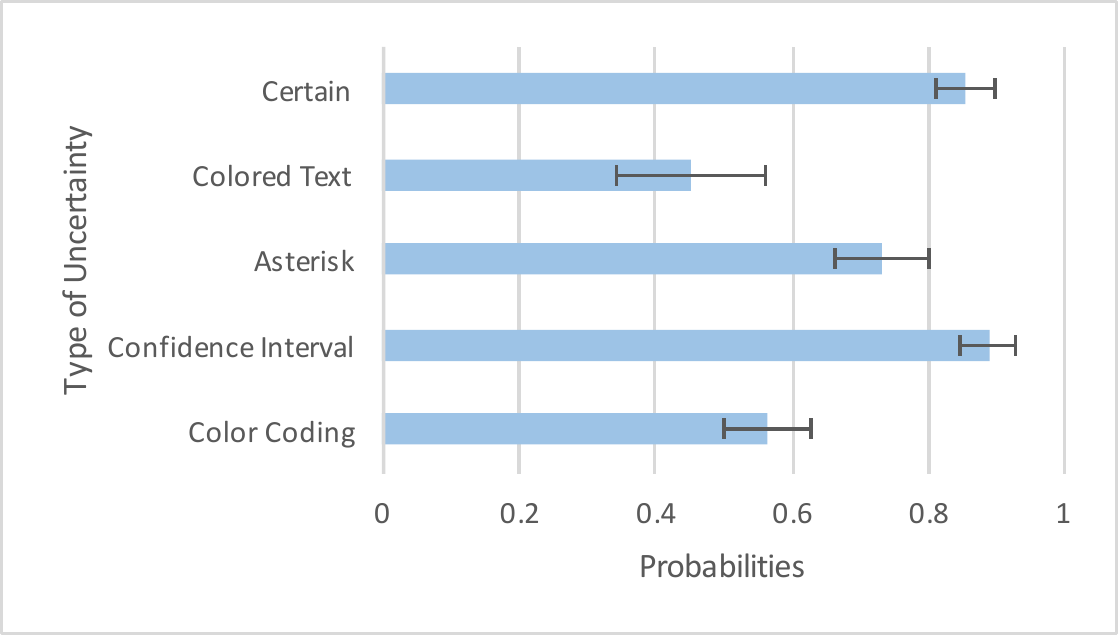}
\caption{\textbf{Probability of the user's selection agreeing with the BestOf3 ranking.}}
\label{fig:bestof3}
\end{figure}

\tinysection{Effectiveness}
The data presented to users was carefully selected to have two properties:  First, each dataset was selected to elicit a specific ordering, regardless of whether participants made their choice based on the best two ratings or based on the average of all three ratings.  We term this ranking order \textbf{BestOf3}.  Second, uncertainty annotations were applied to specific cells of the table specifically to create ambiguity.  As a consequence, we would expect users who chose to disregard uncertain data entirely to pick orderings effectively at random relative to \textbf{BestOf3}.  

In short, if a representation of uncertainty is effective at communicating uncertainty, we would expect to see a more random product ranking.  In the confidence interval representation --- where bounds were not wide enough to prompt a significant level of ambiguity --- we would expect to see ranking close to \textbf{BestOf3}.  

Figure~\ref{fig:bestof3} summarizes our results, showing the probability of agreement between the participant-selected ordering and the \textbf{BestOf3} ordering.  Standard deviations are computed under the assumption that agreement with \textbf{BestOf3} follows a Beta-Bernoulli distribution.  A 16.7\% agreement would indicate a purely random ordering.  The `certain', deterministic baseline shows a consistent, roughly 85\% agreement with \textbf{BestOf3}, and as predicted, so does the confidence interval presentation (89\%).  Both colored text and color coding significantly altered participant behavior (45\% and 56\% agreement with \textbf{BestOf3}).  Asterisks were not as effective at altering participant behavior (73\% agreement).  This is consistent with colored text and color coding signaling significant errors, while asterisks signal caveats or minor considerations on the values presented.

\begin{figure*}
\centering
  \begin{subfigure}[t]{0.49\textwidth}
    \centering
    \includegraphics[width=\textwidth]{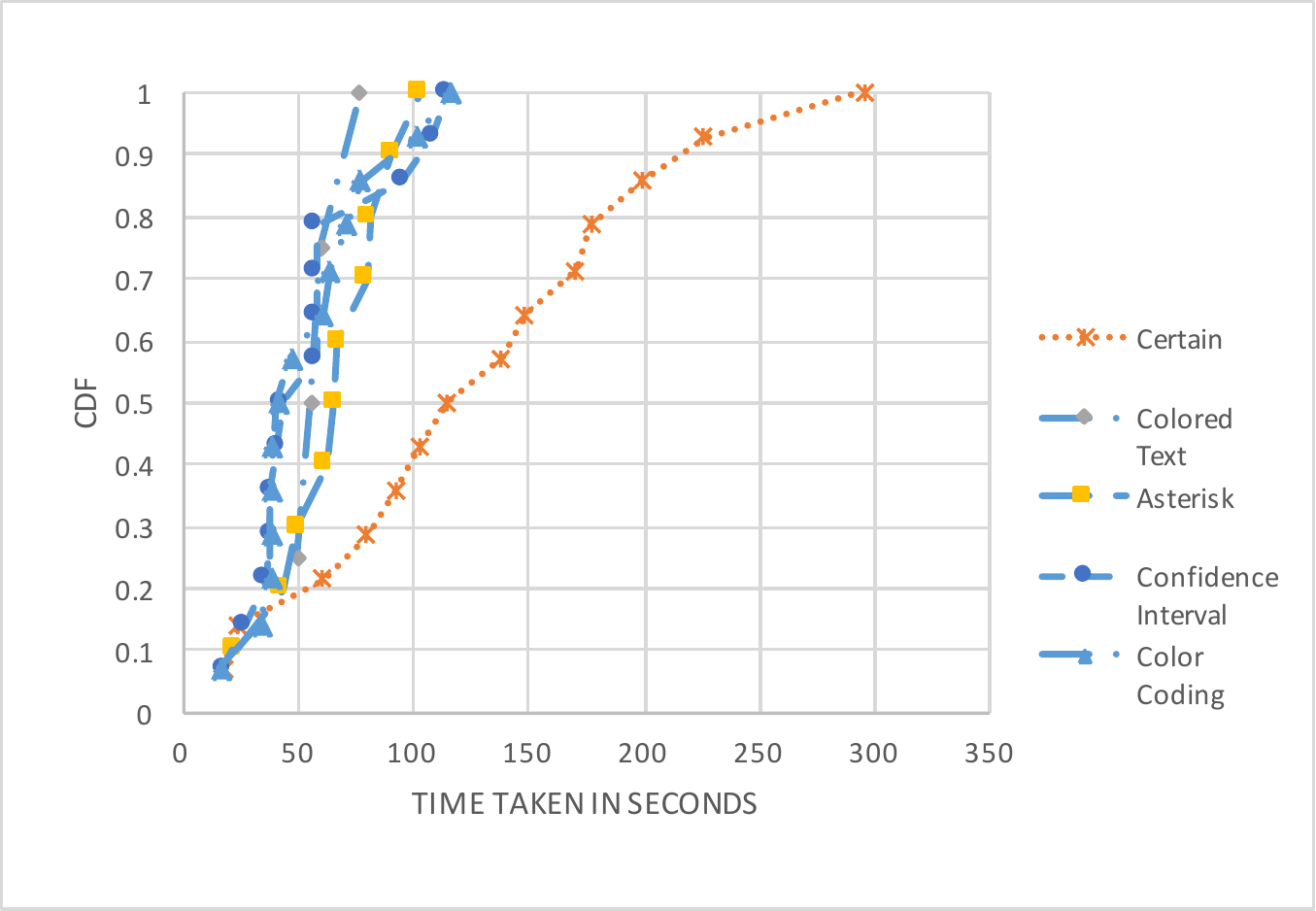}
    \caption{First Round}
    \label{fig:time:first}
  \end{subfigure}
  \begin{subfigure}[t]{0.49\textwidth}
    \centering
    \includegraphics[width=\textwidth]{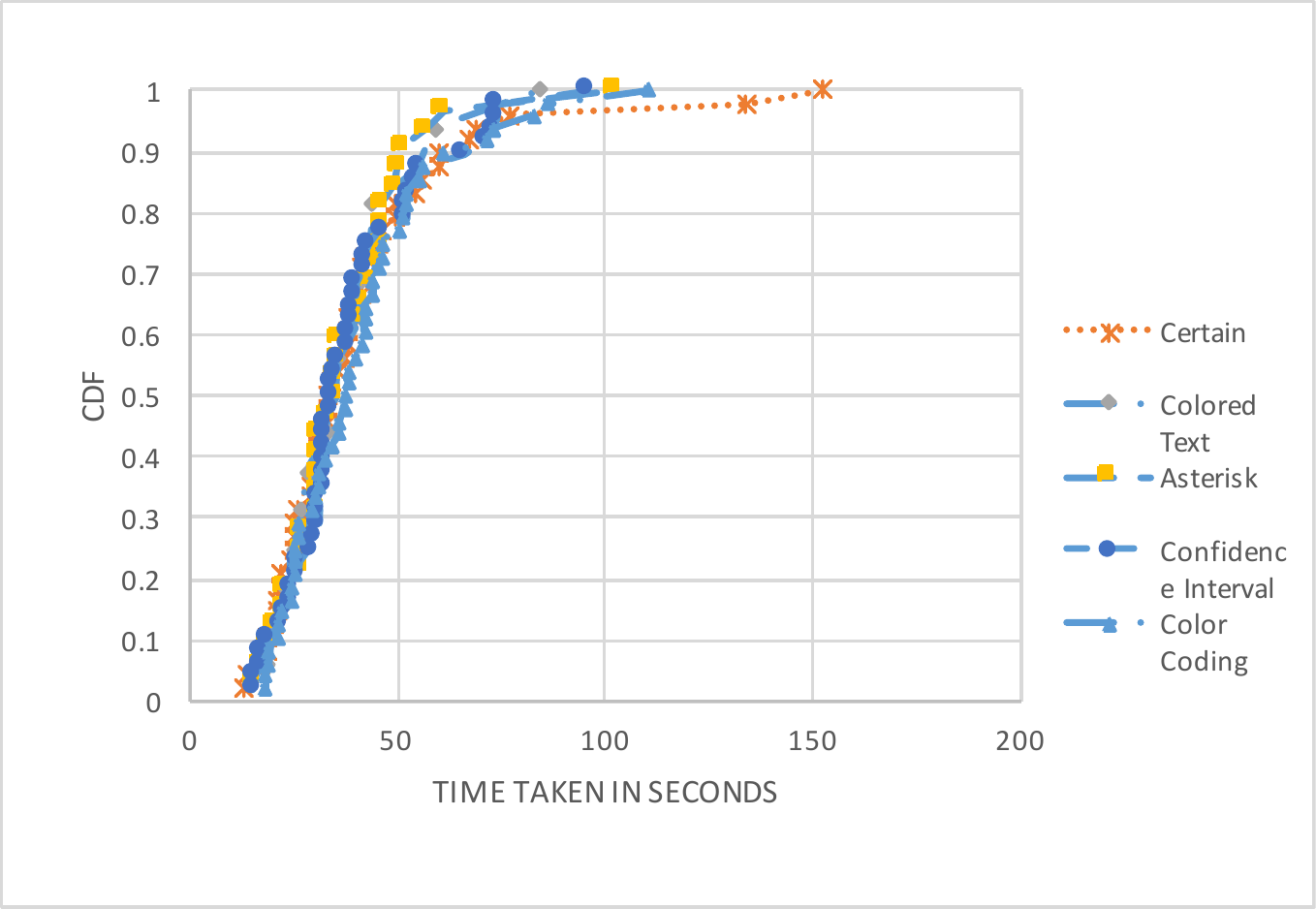}
    \caption{Second to Fifth Rounds}
    \label{fig:time:rest}
  \end{subfigure}
\vspace{-2mm}
\caption{\textbf{Time taken per form of uncertainty.  Graphs show cumulative distributions per-trial.}}
\label{fig:time}
\vspace*{-4mm}
\end{figure*}

\tinysection{Efficiency}
We measure time taken for each form of uncertainty as a proxy for cognitive burden.  Figure~\ref{fig:time} illustrates time taken by users to complete each individual ranking task.  We distinguish between the first round, where participants initially encounter the task and representation, from subsequent rounds where they are already familiar with the task.  As seen in Figure~\ref{fig:time:first}, participants spent significantly more time familiarizing themselves with the overall ranking task than with any of the specific representations of uncertainty.  Furthermore, time taken per representation was relatively consistent across all forms of uncertainty; The slowest two in Figure~\ref{fig:time:rest} trials were both deterministic.  

\section{Discussion}
Participants were encouraged to verbalize their thought process.  Based on this feedback, we were also also able to make several qualitative observations.
In general participants considered consistency in the rating sources and products as a secondary source of feedback about data quality.
For example, if Source 1 had low ratings for all three products, then some participants were more likely to discard it as uninformative and base their rating solely on the other two sources.
If the range of ratings for a product was wide (4.5, 2, 1) then the product was considered unreliable by a few participants. Most of the participants explicitly stated that they were choosing based on the best two of, or the average of the three ratings.

Approximately half of the participants conveyed a strong negative emotional reaction to the color coding representation.  Reactions ranged from participants who expressed a feeling of negative surprise on first seeing the value to participants indicating that the red boxes made them scared.  By comparison, several participants suggested feelings of comfort associated with the additional information that the confidence interval supplied. 

In addition to strong negative emotional responses, most participants indicated that they were ignoring values with a red background, except as a tiebreaker.  This was true even for several participants who did not react in the same way toward the red text or asterisk representations. 

Most participants exhibited risk-averse behavior.  Given two similar choices, many participants stated a preference for products with more consistency in their ratings, as well as for products that did not include uncertain ratings.  
A frequent exception to this pattern was cases where uncertain values appeared at the low end of the rating spectrum --- several participants indicated that the true value of a low, uncertain rating could only be greater than the value being shown.  

In several instances, participants requested additional information, most frequently with the asterisk representation.  It is possible that this is an artifact of the experimental protocol; The asterisk was the first form of uncertainty that many participants encountered.  However, based on our efficacy analysis, it may also be the case that participants assumed that this representation signaled less significant errors.  In future trials, we will use a random trial order and evaluate whether some representations are better at prompting users to seek out additional information.

For confidence bounds, users appeared to react to the presented uncertainty in one of two ways.  One group appeared to first evaluate whether the uncertainty would make a significant impact on their deterministic ranking strategy (best 2 of 3 or average).  The other group adopted a pessimistic view and plugged the lower bound into their deterministic strategy as a worst-case.  For the experimental protocol used, both strategies typically resulted in the same outcome.

\section{Conclusions and Future Work}
Data quality is becoming an increasingly painful challenge to scale.  As a result of issues ranging from low-quality source data~\cite{Kandel:2011:WIV:1978942.1979444,Yang:2015:LOA:2824032.2824055,DBLP:journals/corr/NandiYKGFLG16} to time-constrained execution~\cite{Hellerstein:1997:OA:253262.253291,Kennedy:2011:JEO:1989323.1989410}, the future is clear: Before long, imprecise database query results will be common.  It is thus imperative that we learn how to communicate uncertainty in results effectively and efficiently.  We presented our initial exploration of this space: a user study that examined four approaches to presenting attribute-level uncertainty.  We plan to continue these efforts by exploring (1) other types of uncertainty in relational data (row-level and open-world), (2) qualitative feedback such as explanations \cite{Yang:2015:LOA:2824032.2824055}, (3) giving the user mechanisms to dynamically control the level and complexity of uncertainty representation being shown, and (4) incorporating our findings into the Mimir on-demand curation system~\cite{Yang:2015:LOA:2824032.2824055,DBLP:journals/corr/NandiYKGFLG16}.

\medskip

\noindent \textbf{Acknowledgements}
\textit{
This work was supported in part by a gift from Oracle.   Opinions, findings and conclusions or recommendations expressed in this material are those of the authors and do not necessarily reflect the views of Oracle.
}

\bibliographystyle{abbrv}
\bibliography{references}
\end{document}